\documentclass{ws-rv9x6}  
\usepackage{amsmath}
\newtheorem{define}{Definition}

\begin{document}

\setcounter{chapter}{0}

\chapter{DYNAMICAL SYSTEMS, STABILITY, AND CHAOS}

\markboth{R. Ball and P. Holmes}{Dynamical systems, stability, and chaos}

\author{Rowena Ball}

\address{Mathematical Sciences Institute
and Department of Theoretical Physics \\The Australian National University,               
Canberra, Australia.\\
E-mail: Rowena.Ball@anu.edu.au}

\author{Philip Holmes}

\address{  Department of Mechanical and Aerospace Engineering
and Program in Applied and Computational Mathematics,
    Princeton University, NJ 08544, USA.}

\begin{abstract}
In this expository and resources chapter we review selected aspects of the mathematics of dynamical systems, stability, and chaos,  within a  historical framework that draws together two threads of its early development: celestial mechanics and control theory, and focussing on  qualitative theory. From this perspective we show how concepts of stability enable us to classify dynamical equations and their solutions and connect the key issues of nonlinearity, bifurcation, control, and uncertainty that are common to time-dependent problems in natural and engineered systems. We discuss stability and bifurcations in three simple model problems, and conclude with a survey of recent extensions of stability theory to complex networks. 

\end{abstract}
\section{Introduction}   \label{introduction}

Deep in the heart of northern England, on the banks of a river near a village at the edge of the Lancashire Pennines, there is a fine brick building dating from the late nineteenth century. Here dwell two stout, well-preserved old ladies named Victoria and Alexandra. They will never invite you in for tea though, for the building is the Ellenroad Mill Engine House and the two Ladies are a giant, twin compound steam engine operating in tandem, originally built in 1892. On weekends willing teams of overalled maids and butlers oil and polish the Ladies and fire up the old Lancashire boiler that delivers the steam to their cylinders to move the pistons that drive the giant, 80-ton flywheel. 

\begin{figure}
\hbox{
\begin{minipage}{0.7\textwidth}
\includegraphics[scale=1]{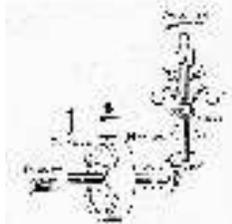}
\end{minipage}
\begin{minipage}{0.3\textwidth}
\caption{\label{governor-sketch} The centrifugal flyball~governor (after Pontryagin (1962)$^{20}$). See Equations \ref{vysh-governor} and accompanying text in section \ref{sec3} for definitions of the labels.  }
\end{minipage}
}
\end{figure}

The speed of the engines is controlled by a centrifugal governor\footnote{The Greek word for governor is \textit{kubernetes}, from which the mathematician Norbert Wiener (1894--1964) coined the term \textit{cybernetics} as a name for the collective field of automated control and information theory.}, and the motions of this device, occurring on time and spatial scales that can be appreciated by the human visual cortex, 
are fascinating to watch. Originally patented by James Watt in 1789, the centrifugal steam engine governor is the most celebrated prototype example of a self-regulating feedback mechanism. The device consists of two steel balls hinged on a rotating 
shaft which is spun from a belt or gears connected to the flywheel, Figure \ref{governor-sketch}. In stable operation, as the speed of the engine increases the inertia of the flyballs swings the arms outwards, contracting the aperture of a valve which controls the speed of rotation by restricting 
the steam supply. If the engine lags due to an additional, imprecisely known, load (in the mill this might have been another loom connected up to the engine by a belt drive) the flyballs are lowered and the valve opens, increasing the steam supply to compensate. Thus the design of the governor cleverly uses the disturbance itself, or deviation from set-point or desired performance, to actuate the restoring force.  

In certain operating regimes the motions of the governor may lose stability, becoming oscillatory and spasmodic, amplifying the effect of the disturbance and thwarting control of the engine. Nineteenth century engineers called this unstable behaviour \textit{hunting} and devoted much effort to improving the design of centrifugal governors. James Clarke Maxwell was the first to formulate and analyse the stability of the equations of motion of the governor, explaining the onset of hunting behaviour in mathematical terms~\cite{Maxwell:1868,Fuller:1976b}, followed (independently) by Vyshnegradskii\cite{Vyshnegradskii:1876}. We analyse 
Vyshnegradskii's 
equations for the governor's motion in section \ref{sec3}, as an exemplary three dimensional stability problem. 

The self-correcting centrifugal governor is a simple feedback control system because the changes in velocity are fed back to the steam valve. Its widespread adoption during the 18th and 19th centuries dramatically transformed the steam-driven textile mills, the mining industry, and locomotion.  (In 1868, the year Maxwell published ``On Governors''\cite{Maxwell:1868} there were an estimated 75,000 Watt governors in England alone\cite{Denny:2002}.) 
Without this device the incipient industrial revolution could not have progressed,  because steam engines lacking self-control would have remained hopelessly inefficient, monstrous, contraptions, requiring more than the labour that they replaced to control them. 

Watt's iconic governor also embodies a radical change in the philosophy  of science. For several hundred years the mechanical clock, with its precise gears and necessity for human intervention to rewind it or correct error and its complete absence of closed-loop feedback, had been the dominant motif in scientific culture. 
In a common metaphor, the universe was created and ordered by God the Clockmaker. Isaac Newton had no doubt that God had initiated the celestial mechanics of the motions of the planets and intervened when necessary to keep His creation perfectly adjusted and on track\cite{Christianson:1984,Peterson:1993}. The clockwork view was also deeply satisfying to Laplace, one of the most influential mathematicians of the eighteenth and early nineteenth centuries. Stability theory was 
developed some two centuries and more after Newton published his \textit{Principia} (1687), so he could not have known
that the planetary orbits may be what Poincar\'e  called Poisson stable\cite{Poincare:1890,Barrowgreen:1997} (small perturbations are self-correcting) --- or they may be chaotic\footnote{A fact which might cause you some queasiness to learn. Fear not  --- it is believed that chaotic motions were important in the early evolution of the solar system\cite{Morbidelli:2005}, and a slow chaotic drift may be noticeable a few billion years hence\cite{Sussman:1992,Peterson:1993}. What luck for us!}.

As concepts of feedback and stability were developed rigorously and applied in the late nineteenth and early twentieth centuries, Divine open-loop control began to wane and there came a growing awareness of systems as dynamical entities that can regulate their own destiny and internally convert uncertain inputs into stable outputs.  The technological advances in transport, power, and communications made possible by feedback control and applied stability theory are agents of change, the vectors of liberty, liberalism, and literacy
 in societies, themselves enabling the blossoming and seeding of more 
sophisticated ideas of feedback and stability in complex environmental, socio-economic, and biological
 systems.  \emph{Now,} due to stability theory and feedback control, we may contemplate ``the fundamental interconnectedness of all things''\cite{Adams:1987}, but back \emph{then,} in the clockwork days,
people could not. It is surely no coincidence that totalitarian governments favour clockwork metaphors. 

Today, we are so comfortable with the concept of feedback control inducing stable dynamics that we barely notice how it permeates most aspects of our lives.

\subsection*{}
Control theory, then, is a major strand in the development of modern nonlinear dynamics, but it is not the first. The centrifugal governor also transformed the practice of astronomy, in that it enabled fine control of telescope drives and vastly improved quantitative observations, and it is this earlier force (already alluded to above in mention of Newton's and Laplace's work) 
in the development of dynamical systems and stability theory
 --- celestial mechanics --- on which we now focus attention. 
The next stage of our nonlinear dynamics odyssey takes us from the post-industrial north of England to the miraculously intact (given the destructions of WWII) medieval city of Regensburg in Germany, to an older, humbler but no less important building than that which houses the Ladies, the Kepler museum. In addition to celebrating the life and work of Johannes Kepler (1571--1630) the museum houses priceless manuscripts, letters, publications, and astronomical instruments and interpretive exhibits that tell a lively and inspiring story, that of the development of celestial mechanics from Galileo to its culmination, in analytic terms,  in the work of Poincar\'e.

\begin{figure}
\centerline{\hbox{
\begin{minipage}{0.4\textwidth}
\includegraphics[scale=1.2]{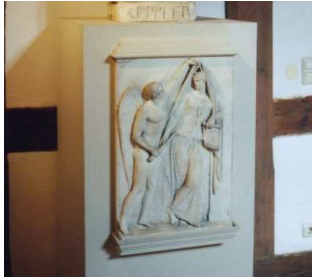}
\end{minipage}
\begin{minipage}{0.4\textwidth}
\caption{\label{kepler}Johannes Kepler is given cheek by his Muse after two long centuries of ellipses and clockworks. }
\end{minipage}
}}
\end{figure}

An exhibit from the 18th century, an exquisitely engineered  brass orrery, or clockwork model of the solar system,  in its detail and precision expresses the satisfaction and confidence  of the clockwork aficionados of the Age of Enlightenment. But a nearby exhibit expresses, rather presciently, the need for a new metaphor for scientific endeavour and achievement. It is an early 19th century relief in which Kepler unveils the face of Urania, the Muse of astronomy, whereupon she insouciantly hands him a telescope and a scroll inscribed with his own laws, as if to say: ``Hmm\ldots not a bad job; now take these back and do some more work then tell me why your elliptical orbits are non-generic''. (See Figure~\ref{kepler}.)
And in fact the one-dimensional Kepler ellipse can be transformed into   a harmonic oscillator with Hamiltonian\cite{Cvitanovic:2005} 
\begin{equation}\label{kepler-ellipse}
H(Q,P)=\frac{1}{8}P^2 - EQ^2 .
\end{equation}

Despite Laplace's confidence the problem of the stability of the solar system refused to go away, but instead took on a central role in the preoccupations of mathematicians, physicists, astronomers, and navigators post-Newton. It was by no means clear, even to Newton, that Newton's law was sufficient to describe the motions of three or more celestial bodies under mutual gravitational attraction. The problem also refused to be solved, in the sense  of what was accepted as a ``solution'' during the latter 18th century and first half of the 19th century, i.e., analytically in terms of elementary or previously-known special functions.  

Progress was made in the mid-1800s in improving series approximations but, not surprisingly, the hydra of nonconvergence soon raised one after the other of its ugly (of course!) heads.  By 1885, when it was chosen by Weierstrass as one of four problems in the mathematics competition sponsored by King Oscar II of Sweden, the $n$-body problem had achieved notoriety for its recalcitrance --- but  in doing so it had also driven many of the seminal advances in mathematics and produced many of the greatest mathematicians of the 19th century. 

The first problem in King Oscar's competition was to show that the  solar system as modeled by Newton's equations is stable. In his  (corrected) entry\cite{Poincare:1890} Poincar\'e 
invented or substantially extended integral invariants, characteristic exponents, and \textbf{Poincar\'e maps} (obviously), invented and proved the \textbf{recurrence theorem}, proved the nonexistence of uniform first integrals of the three body problem, other than the known ones, discovered \textbf{asymptotic solutions} and \textbf{homoclinic points},  and wrote the first ever description of chaotic motion --- in short, founded and developed  the entire subject of  geometric and qualitative analysis. Then he concluded by saying he regarded his work as only a preliminary survey from which he hoped future progress would result. 

Poincar\'e's ``preliminary survey'' is still inspiring new  mathematics and applications, but during the 20th century the collective dynamic of dynamical systems development was highly nonlinear. Homoclinic points and homoclinic chaos were partially treated by the American mathematician George Birkhoff (1884--1944) --- he obtained rigorous results on the existence of periodic orbits \textit{near} a homoclinic orbit ---  and by Cartwright and Littlewood in their study of Van der Pol's (non-Hamiltonian) equation\cite{Cartwright:1945},
$$\ddot{y}-k(1-y^2)\dot{y}+y=b\lambda k\cos(\lambda t _\alpha) .$$
Cartwright and Littlewood stated numerous ``bizarre'' properties of solutions of this differential equation, implying the existence of an invariant Cantor set, but their very concise paper was not easy to penetrate, and their results remained largely unknown until Levinson\cite{Levinson:1949} pointed them out to Stephen Smale.

During the 1960s and 1970s Smale's representation of homoclinic chaos in terms of symbolic dynamics and the horseshoe map\cite{Smale:1980} stimulated renewed interest in dynamical systems (although we have skipped a lot of mathematical history here, most notably KAM theory).  Happily, this coincided with the advent of desktop digital computers subject to Moore's law. Since the 1980s  improvements in processor speed have both driven and been driven by the use of computational simulations of dynamical systems as virtual experiments, and inspired advances in fields such as network stability, numerical instabilities, and turbulence. Essentially these advances are sophisticated and technologically facilitated applications of Poincar\'e's and Lyapunov's stability theory, and in the next section we present the basics and some working definitions. 

It is somewhat ironic that improvements in processor speed have also led to renewed interest in low dimensional dynamical systems, which usually only require small-time computing and are at least partially amenable to rigorous stability analysis. For large dynamical systems usually mean turbulent ones, and computation is, in essence, the notorious ``problem of turbulence''. In a turbulent flow energy is distributed among wavenumbers that range over perhaps seven orders of magnitude (for, say, a tokamak) to twelve orders of magnitude (for a really huge system, say a supernova). To simulate a turbulent flow in the computer it is necessary to resolve all relevant scales of motion in three dimensions. It is a fair estimate\cite{Oran:2006} that such calculations would take 400 years at today's processor speeds, therefore a faster way to do them would be to rely on Moore's law and wait only 20 years until computers are speedy enough. 

Many of us in the turbulence business have realized that while we are waiting we can, more expediently, apply reduced dynamical systems methods to the  problem, such as Karhunen-Lo\'eve (KL) decomposition\footnote{Also known by the aliases proper orthogonal or singular value
decomposition, principal component analysis, and empirical eigenfunction analysis.}, to distill out a much-reduced, but nevertheless sophisticated, approximation to the dynamics and spatial structure of a turbulent  flow\cite{Holmes:1996}. To introduce KL decomposition, we imagine a fractional distillation tower for which the feedstock is not crude oil but a high Reynolds number flow. Then instead of a natural distribution over hydrocarbon molecular weights we have an energy distribution over scales of motion. We know, in principle, how hydrocarbons are separated in the still according to their boiling points (even if we do not work at an oil refinery), but what properties may we exploit to separate and re-form the energy components of a turbulent flow? Our turbulence refinery does not define the skyline of a seamy port city in complicated chiaroscuro, but exists more conveniently in constrained fluid flow experiments or as direct numerical simulations of the Navier-Stokes equations \textit{in silico}. The KL transform operates on data to yield eigenfunctions that capture in decreasing order most of the kinetic energy of the system, so it is especially useful for highly self-structured flows.\\[3mm]
\centerline{
\includegraphics[scale=0.5]{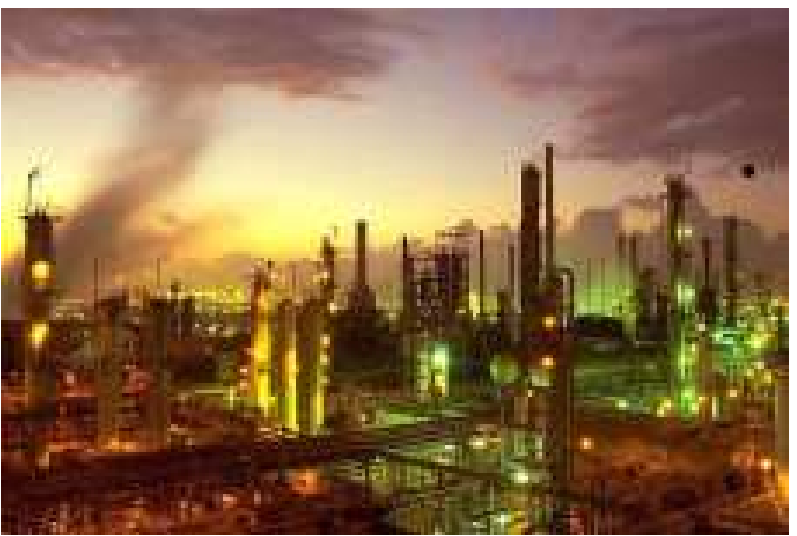}
}

\section{To understand stability is to understand dynamics}

Very few dynamical systems have known, exact solutions. For the vast majority it cannot even be proved that general solutions exist. Stability theory is quite indifferent to such issues; instead it tells us how families of solutions would behave, assuming they \textit{do} exist. Loosely we understand  stability to mean that a solution does not run away, or to refer to the resilience of a solution to changes in initial conditions or to changes to the equation that generates it. Stability  is a qualitative property of dynamical equations and their solutions. 

For practical applications stability analysis allows us to say whether a given system configuration will exhibit runaway dynamics (catastrophic failure) or return to a stable quasi-equilibrium, limit cycle, or other attractor, 
in  response to perturbation. We have indicated in section \ref{introduction} above how the issue of stability of the planetary orbits drove the development of celestial mechanics, but stability is equally important in control
theory --- from a design and operational point of view it could be said that
control \emph{is} applied stability. It is a grave issue because, as we show in section \ref{sec3},  feedback can result in systems that fail due to instabilities, as well as create  ones which maintain homeostasis. Thermal explosions, ecological ``arms races'',  and economic depressions are all more-or-less disastrous consequences
of unstable feedback dynamics. A big stability question that occupies many
scientists today concerns the long-term stability of the world's climate in response to the enhanced greenhouse effect;  questions related to stability of other complex systems will be explored in section \ref{sec6}.

In this section we give precise mathematical expression to  these concepts of stability, for later reference.  For more detail and discussion  the reader is referred to the article in Scholarpedia curated by Holmes and Shea-Brown\cite{Holmes:2006}. 

Consider the general dynamical system in vector form
\begin{equation}\label{general}
\dot{\mathbf{x}} = \mathbf{f}(t,\mathbf{x}), 
\end{equation}
where $f^i(t,\mathbf{x})$ and the derivatives $\partial f^i(t,\mathbf{x})/\partial x^j$ are defined and continuous on a domain $\Gamma$ of the space of $t, \mathbf{x}$. 
Let $\gamma_t(\mathbf{x}) = \mathbf{x}(t)$ with the initial value $\mathbf{x}(0)=\mathbf{x}$. Then, the (forward) orbit is the set of all values that this trajectory obtains: $\gamma(\mathbf{x}) = \left\{ \gamma_t(\mathbf{x}) | t \ge 0 \right\}$. 

\begin{define}{} 
Two orbits $\gamma(\textbf{x})$ and $\gamma(\hat{\mathbf{{x}}})$ are $\epsilon$-close if there is a reparameterization of time (a smooth, monotonic function) $\hat{t}(t)$ such that $    | \gamma_t(\mathbf{x}) - \gamma_{\hat{t}(t)}(\hat{\mathbf{{x}}}) | < \epsilon \mbox{ for all } t \ge 0$. 
\end{define}

\begin{define} \textbf{Orbital or generalized Lyapunov stability}.  $\gamma(\textbf{x})$ is orbitally stable if, for any $\epsilon>0$, there is a neighbourhood $V$ of $\mathbf{x}$ so that, for all $\hat{\mathbf{x}}$ in $V$, $\gamma(\mathbf{x})$ and $\gamma(\hat{\mathbf{x}})$ are $\epsilon$-close.
\end{define}

\begin{figure}\centerline{
\includegraphics[scale=1]{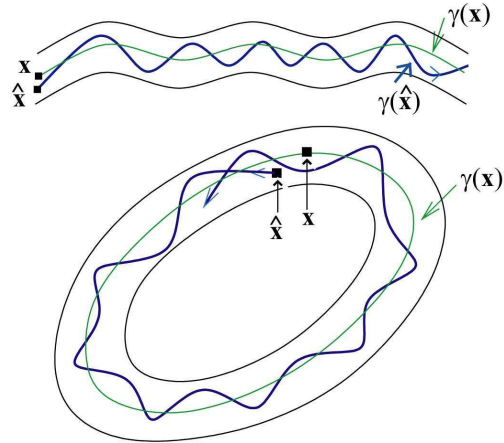}}
\caption{\label{orbital-stability} The orbit $\gamma(\mathbf{x})$ is orbitally stable. The black lines indicate the boundary of an $\epsilon$-neighborhood of ${\gamma}(\mathbf{x})$.}
\end{figure}

\begin{define} \textbf{Generalised asymptotic  stability}. If additionally $V$ may be chosen so that, for all $\hat{\mathbf{x}} \in V$, there exists a constant $\tau(\hat{\mathbf{x}})$ so that $| \gamma_t(\mathbf{x}) - \gamma_{t-\tau(\hat{\mathbf{x}})}(\hat{\mathbf{x}}) | \rightarrow 0 \mbox{ as } t \rightarrow \infty$  
then $\gamma_t(\mathbf{x})$ is asymptotically stable. 
\end{define}
These general definitions of Lyapunov stability and asymptotic stability are indifferent to the choice of initial values $t_0,\mathbf{x}(0)$. Lyapunov stability is intimated in Figure \ref{orbital-stability}, which sketches a segment of an orbit ${\gamma}(\mathbf{x})$ and a segment of a neighbouring orbit ${\gamma}(\hat{\mathbf{x}})$, in periodic and non-periodic cases. 

\begin{figure}\centerline{
\includegraphics[scale=1]{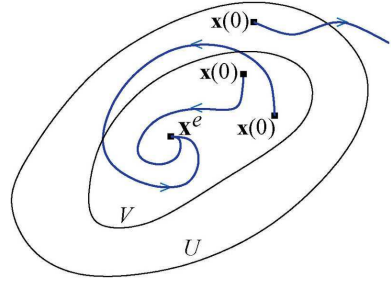}}
\caption{\label{a-equil-stability} An asymptotically stable equilibrium is also called a sink. }
\end{figure}

In the particular case where the system (\ref{general}) is autonomous and the solution is an equilibrium $\mathbf{x}^e$ we have the following specifications:
\begin{define}
\textbf{ Lyapunov stability of equilibria}.
$\mathbf{x}^e$ is a stable equilibrium if for every neighborhood $U$ of $\mathbf{x}^e$ there is a neighborhood $V \subseteq U$ of $\mathbf{x}^e$ such that every solution $\mathbf{x}(t)$ starting in $V (\mathbf{x}(0) \in V)$ remains in $U$ for all $t \geq 0$. Notice that $\mathbf{x}(t)$ need not approach $\mathbf{x}^e$. Lyapunov stability means that when all orbits starting from a small neighbourhood of a solution remain forever in a small neighborhood of that solution the motion is stable, otherwise it is unstable. 
If $\mathbf{x}^e$ is not stable, it is unstable. 
\end{define}
\begin{define}
\textbf{ Asymptotic stability of equilibria}.
An equilibrium $\mathbf{x}^e$ is asymptotically stable if it is Lyapunov stable and additionally $V$ can be chosen so that $|\mathbf{x}(t) - \mathbf{x}^e | \to 0$ as $t \to \infty$ for all $\mathbf{x}(0) \in V$. An asymptotically stable equilibrium (stationary state) and its local environment is sketched in Figure \ref{a-equil-stability}.
\end{define}

It is all very well to settle the stability properties of a solution, but what then? If, as is usually the case, we are studying Eq. \ref{general} as a model for coupled physical motions or a system of rate processes, and therefore necessarily imperfect, we also need information about how those properties fare under perturbations to the model, or \textit{structural stability}. The question usually goes something like this: When are sufficiently small perturbations of a dynamical system equivalent to the original unperturbed dynamical system? And if a system is not structurally stable, how may one unfold it until it is? And what (new mathematics, physics) do the unfoldings reveal? The concept of structural stability has yielded a rich taxonomy of bifurcations and of different classes of vector fields.  Structural stability is thus fundamentally a classification science, a binomial key of the type that has been used in biology since the method was devised by the Swedish botanist Linnaeus (1707–-1778).   It is more distracting than useful to  define structural stability rigorously at this stage (although authoritative  definitions can be found in the literature, e.g., Hirsch and Smale (1974)\cite{Hirsch:1974}); instead, we shall illustrate some of the concepts in section \ref{sec4} in relation to a perturbed simple pendulum as a simplified surrogate for the restricted three body problem.

\section{Governor equations of motion: a simple case study}\label{sec3}
Now that we have some background and theory resources to draw on, let us carry out  a stability analysis of the centrifugal governor. This analysis is all the more important for being elementary because it introduces many of the key concepts of dynamical systems theory in a setting that is understandable to non-mathematician physical scientists and engineers and also sets the scene for the  more complicated  motions we describe in sections \ref{sec4} and \ref{sec5}.  
Vyshnegradskii's equations of motion for the flyball governor  sketched in Figure  \ref{governor-sketch} were given as a 3-dimensional, autonomous, first-order dynamical system by Pontryagin (1962)\cite{Pontryagin:1962}: 
\begin{equation}\label{vysh-governor}
\begin{aligned}
\frac{d\varphi}{dt}&= \psi\\
\frac{d\psi}{dt}&= n^2\omega^2\sin\varphi\cos\varphi -
                   g\sin\varphi - \frac{b}{m}\psi\\
\frac{d\omega}{dt}&= \frac{k}{J}\cos\phi - \frac{F}{J},
\end{aligned}
\end{equation}
where $\varphi$ is the angle between the spindle $S$ and the flyball arms $L$, $\omega$ is the rotational velocity of the flywheel, the transmission ratio $n=\theta/\omega$, $\theta$ is the angular velocity of $S$, $g$ is the gravitational acceleration, $m$ is the flyball mass, $J$ is the moment of inertia of the flywheel, $F$ represents the net load on the engine, $k>0$ is a constant, and $b$ is a frictional coefficient. The length of the arms $L$ is taken as unity. 
For a given load $F$ the engine speed and  fly-ball angle are required to remain constant,  
and the unique steady state or equilibrium coordinates are easily found as $\psi_0=0$, $\cos\varphi_0=F/k$, $n^2\omega_0^2=g/\cos\varphi_0$. So far, so dull.  

Dull, too, are the designers of engines, according to Maxwell. In his treatment of the governor problem, which was more general than that of Vyshnegradskii, he wrote:   ``The actual motions corresponding to these impossible roots are not generally taken notice of by the inventors of such machines, who naturally confine their attention to the way in which it is \textit{designed} to act; and this is generally expressed by the real root of the equation.'' The impossible roots he referred to are the complex roots of the characteristic equation obtained from the linearized equations of motion. Maxwell and Vyshnegradskii both used this method to investigate the mathematical stability of the engine-governor dynamical system and relate the results closely to observed misbehaviours of the physical system. Their linear stability analyses provide criteria for which the system returns to its equilibrium engine speed $\omega_0$ and flyball angle $\varphi_0$ when subjected to a small perturbation.  Let us represent the perturbed system by setting 
$$ \varphi=\varphi_0+\delta\varphi,\quad\psi=\psi_0+\delta\psi,\quad\omega=\omega_0+\delta\omega,$$ with 
$|\delta\varphi|$, $|\delta\psi|$, $|\delta\omega| \ll 1$,
and recasting equations (\ref{vysh-governor}) as
\begin{equation}\label{perturbed-vysh-governor}
\begin{aligned}
\frac{d}{dt}\delta\varphi&= \delta\psi\\
\frac{d}{dt}\delta\psi&= -\frac{g\sin^2\varphi_0}{\cos\varphi_0}\delta\varphi
                         - \frac{b}{m}\delta\psi 
                           + \frac{2g\sin\varphi_0}{\omega_0}\delta\omega\\
\frac{d}{dt}\delta\omega&= -\frac{k}{J}\sin\varphi_0\delta\varphi,
\end{aligned}
\end{equation}
where we have neglected terms that are quadratic in the small perturbations $\delta\varphi$, $\delta\psi$, and $\delta\omega$. Equations (\ref{perturbed-vysh-governor}) are a linear system with constant coefficients that may be written succintly in matrix form
\begin{equation}\label{linear}
\mathbf{\dot{x}}=\mathbf{A}\mathbf{x},
\end{equation}
where 
$$
\mathbf{\dot{x}}=\begin{pmatrix}
      \frac{d}{dt}{\delta\varphi}\\
      \frac{d}{dt}{\delta\psi}\\
      \frac{d}{dt}{\delta\omega}     
     \end{pmatrix},\quad
\mathbf{A}= \begin{pmatrix}
       0&1&0\\
       -\frac{g\sin^2\varphi_0}{\cos\varphi_0} &  - \frac{b}{m} & \frac{2g\sin\varphi_0}{\omega_0}\\
        -\frac{k}{J}\sin\varphi_0 & 0 &0
     \end{pmatrix}, \quad
\mathbf{x}=\begin{pmatrix}
      \delta\varphi\\
      \delta\psi\\
      \delta\omega  
     \end{pmatrix}.
$$
Equation (\ref{linear}) has nontrivial, linearly independent solutions of the form 
\begin{equation}\label{s1}
\mathbf{x} = \mathbf{u}e^{\lambda t}
\end{equation} 
 where the constant components of $\mathbf{u} $ and the constant $\lambda$ may be complex. 
Differentiating (\ref{s1}) with respect to $t$ and substituting 
 in (\ref{linear}) gives the eigenvalue problem
\begin{equation}\label{s2}
\left(\mathbf{A} - \lambda \mathbf{I}\right)\mathbf{u}=0
\end{equation}
where $\mathbf{I}$ is the identity matrix. The requirement that $\mathbf{u}\neq 0$, needed to obtain nontrivial solutions, satisfies (\ref{s2}) if and only if the factor
$\det\left(\mathbf{A}- \lambda \mathbf{I}\right)=0$, 
or 
\begin{equation}\label{determinant}
\begin{vmatrix}
       -\lambda &1&0\\
       -\frac{g\sin^2\varphi_0}{\cos\varphi_0} &  - \frac{b}{m}-\lambda & \frac{2g\sin\varphi_0}{\omega_0}\\
        -\frac{k}{J}sin\varphi_0 & 0 &-\lambda   
\end{vmatrix}
=0. 
\end{equation}
The determinant may be evaluated and equation (\ref{determinant}) expressed in terms of the characteristic polynomial:
\begin{equation}\label{characteristic}
\lambda^3 + \frac{b}{m}\lambda^2 + \frac{g\sin^2\varphi_0}{\cos\varphi_0} \lambda + 
\frac{2gk\sin^2\varphi_0}{J\omega_0}=0. 
\end{equation} 
The roots $\lambda_1,\lambda_2,\lambda_3$ of (\ref{characteristic}) are the eigenvalues of $A$ and the solutions $\mathbf{u}_1,\mathbf{u}_2,\mathbf{u}_3$ of (\ref{s2}) are the corresponding eigenvectors. 
By inspection of equation (\ref{s1}) stability can ensue only if the real eigenvalues, or real parts of complex eigenvalues are negative. From analysis of the characteristic equation (\ref{characteristic}) this condition can be written as  
 \begin{equation}\label{condition} 
 \frac{bJ}{m} \frac{\omega_0}{2F}>1.
\end{equation}

\begin{figure}[ht]
\centerline{
\includegraphics[scale=0.8]{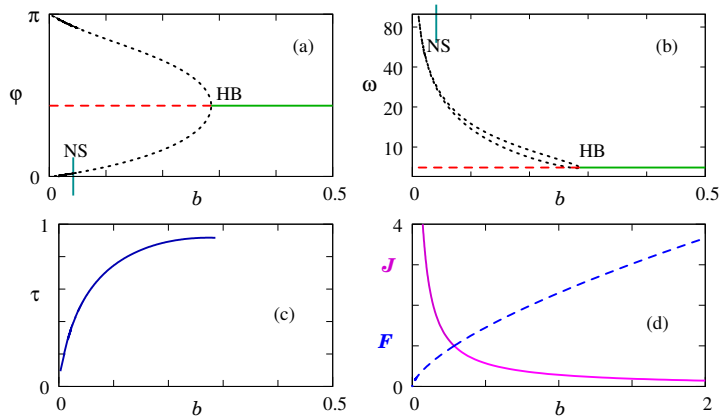}}
\caption{\label{bifurcation-diagrams} Bifurcation diagrams rendered for the variables $\varphi$ (a) and $\omega$ (b), stable equilibria are marked by a solid line, unstable equilibria are marked by a dashed line, HB stands for Hopf bifurcation, NS stands for Neimark-Sacker bifurcation, black dots mark the amplitude envelope of the oscillations. (c) The period $\tau$ of the oscillations decreases with $b$. (d) Continuations at the Hopf bifurcation in the parameters $J$ and $F$.  }
 \end{figure}

Now let us consider the dynamical behaviour of the engine-governor system in the light of  (\ref{condition}) and with the aid of Figure \ref{bifurcation-diagrams}. In (a) and (b) the equilibria and linear stability of equations \ref{vysh-governor} have been computed numerically and plotted as a function of the friction coefficient $b$. This is a bifurcation diagram, where the bifurcation or control parameter $b$ is assumed to be quasistatically variable, rendered in the variables $\varphi$ (a) and $\omega$ (b). We see immediately that stable, steady state operation of the engine-governor system requires frictional dissipation above a critical value. As $b$ is decreased through the \textbf{Hopf bifurcation} point HB the real parts of a pair of conjugate eigenvalues become positive, the equilibrium becomes unstable, and the motion becomes oscillatory. The envelope of the periodic solutions grows as $b$ is decreased further, which is also deduced in the inequality (\ref{condition}):  a decrease in the coefficient of friction can destabilize the system. 

As the bifurcation parameter $b$ is decreased through the marked value with the label NS the stable periodic solution, for which the Floquet multipliers have modulus $< 1$,  undergoes a \textbf{Niemark-Sacker bifurcation}. A conjugate pair of multipliers leaves the unit circle, and a two-dimensional asymptotically stable invariant torus bifurcates from the limit cycle\cite{Niemark:1959,Sacker:1964} \footnote{The discovery of torus bifurcations first by Niemark in the USSR and five years later independently by Sacker in the USA seems to be a classic case of unnecessarily duplicated development of mathematics during the cold war.}. For $b<b_{NS}$ the periodic solutions are unstable but the torus is stable.  The behaviour of the system has become essentially 3-dimensional. 

\subsubsection*{}
In the governor problem we have studied the stability of solutions. In the next section we consider structural stability, in relation to the the restricted three body problem from celestial mechanics. 

\section{The restricted three body problem,  homoclinic chaos, and structural stability}\label{sec4} 

This section assumes a working knowledge of Hamiltonian mechanics from a text book such as Goldstein (1980)\cite{Goldstein:1980} or from undergraduate lecture notes such as Dewar (2001)\cite{Dewar:2001}. Rather than presume to capture the entire content and context of the restricted three body problem within the space of one chapter section we again summarize a small vignette from the panorama, a surrogate for the restricted three body problem. Homoclinic chaos and the associated topics of Poincar\'e maps, symbolic dynamics, and the Smale horseshoe construction, are fleshed out in Guckenheimer and Holmes (1983)\cite{Guck:1983} and Holmes (1990)\cite{Holmes:1990}. 

\begin{figure}
\centerline{
\includegraphics[scale=0.7]{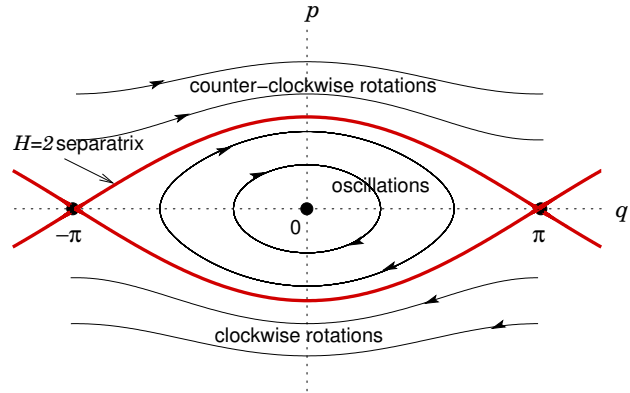}
}
\caption{\label{pendulum} The phase space of the simple pendulum}
\end{figure}

First let us return to Kepler's ellipse, or the two-body problem of Newton, which at the end of section \ref{introduction} we gave in terms of the Hamiltonian for the transformed harmonic oscillator, Eq. \ref{kepler-ellipse}. The well-known simple pendulum is also a harmonic oscillator, with Hamiltonian 
\begin{equation}\label{hpem}
H = p^2/2 + (1-\cos q)
\end{equation}
and equations of motion 
\begin{equation}\label{pem}
\dot{q}=p, \quad \dot{p}= - \sin q.
\end{equation}
The phase portrait of the flow, Figure \ref{pendulum}, 
shows the three families of periodic solutions bounded by the separatrices $H=2$, which are emphasized in Figure~\ref{pendulum}.  The fixed point (or equilibrium) at $(q,p)=(0,0)$ represents the pendulum at rest and that at $(q,p)=(\pm \pi,0)$ represents the upside-down position of the pendulum, keeping in mind that the flat phase portrait should be wrapped around a cylinder of circumference $2\pi$.  Elementary linear analysis tells us that the the fixed point at $(q,p)=(0,0)$ is a centre, with the solution matrix of the linearization having a pair of pure imaginary eigenvalues, and that at $(q,p)=(\pm \pi,0)$ is a hyperbolic (or non-degenerate) saddle point, with the solution matrix of the linearization having having one positive and one negative eigenvalue. 
Each point of the $H=2$ separatrices is  homoclinic, or  asymptotic to to the fixed point $(q,p)=(\pm \pi,0)$  as $t\rightarrow \pm \infty$. In fact the separatrices are simultaneously the stable and unstable manifolds for the saddle point. Thus the phase portrait of the pendulum contains  qualitative information about the global dynamics of the system. 

Now consider the restricted three body problem that featured in Poincar\'e's memoir, in which two massive bodies move in circular orbits on a plane with a third body of negligible mass moving under the resulting gravitational potential. In a rotating frame the system is described by the position coordinates $(q_1,q_2)$ of the third body and the conjugate momenta $(p_1,p_2)$. Poincar\'e studied  the following two degree of freedom Hamiltonian as a proxy for this system:
\begin{equation}\label{proxy}
H(q_1,q_2,p_1,p_2)= -p_2 -p_1^2 +2\mu\sin^2(q_1/2)
+\mu\varepsilon\sin q_1\cos q_2, 
\end{equation}
with corresponding equations of motion 
\begin{equation}
\begin{split}
\dot{q}_1 &= -2p_1, \quad \dot{q}_2 = -1;\\ \dot{p}_1&=-\mu\sin q_1-\mu\varepsilon \cos q_1\cos q_2, \quad \dot{p}_2= \mu\varepsilon\sin q_1\sin q_2. 
\end{split}
\end{equation}
By inversion of Eq. \ref{proxy} we have 
\begin{equation}
p_2=P_h(q_1,p_1;q_2)= h-p_1^2 +2\mu\sin^2(q_1/2)+\mu\varepsilon\sin q_1\cos q_2,
\end{equation}
 from which we can obtain the reduced equations of motion 
\begin{equation}\label{reduced}
q^\prime_1= -\partial P_h/\partial p_1=2p_1,\quad
p_1^\prime = \partial P_h/\partial q_1 = \mu\sin q_1+\mu\varepsilon\cos q_1\cos q_2, 
\end{equation}
where $(\cdot)^\prime$ denotes $d/dq_2)$. 

\begin{figure}
\centerline{
\includegraphics[scale=0.8]{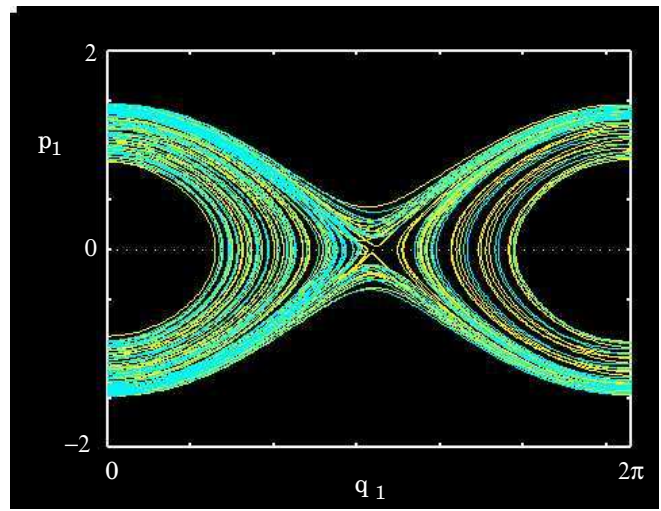}
}
\caption{\label{perturbed-pend} Segments of orbits belonging to the stable (blue) and unstable
(yellow) manifolds of the saddle type periodic orbit of the
periodically perturbed pendulum, Eqs \ref{reduced} with
$\mu=-1$ and $\varepsilon=0.1$. 
 }
\end{figure}

We see that Eqs \ref{reduced} have the form of a periodically forced one degree of freedom system in which the angle variable $q_2$ plays the role of time. 
For $\varepsilon=0$  Eqs \ref{reduced} are isomorphic to those for the simple pendulum, Eqs \ref{pem}, and the phase portrait is that of Figure \ref{pendulum} (to make the origin $(q_1, p_1) = (0, 0)$ a center we set $\mu < 0$). When a time-periodic perturbation is applied to the pendulum the stable and unstable manifolds that form the separatrix level set typically break up, but some homoclinic points may persist and with them small neighbourhoods of initial conditions, which are repeatedly mapped around in the region formerly occupied by the separatrixes. 
Such regions can now fall on both sides of the saddle point so that of two solutions starting near each other, one may find itself on the rotation side and the other on the oscillation side. At each juncture near  the saddle point such solutions must decide which route to take. The global structure of the stable and unstable manifolds rapidly becomes very complicated. Poincar\'e prudently decided that, in this case, a thousand words are worth more than a picture: ``When we try to represent the figure formed by [the stable and unstable manifolds] and their infinitely many intersections, each corresponding to a doubly asymptotic solution, these intersections form a type of trellis, tissue or grid with infinitely fine mesh. Neither of the two curves must ever cross itself again, but it must bend back upon itself in a very complex manner in order to cut across all of the meshes in the grid an infinite number of times.''(Poincar\'e (1899)\cite{Poincare:1899}, quoted in Diacu and Holmes (1996)\cite{Diacu:1996}). We have computed some orbits and rendered the data in Figure \ref{perturbed-pend}, which may or may not help to clarify the issue.  

Thus did Poincar\'e describe homoclinic chaos, after years of careful and productive analysis of the phenomenon. In particular, Poincar\'e obtained the following results:
\begin{itemize}
\item Transverse homoclinic points exist for $\varepsilon\neq 0$. 

A transverse homoclinic orbit occurs when the stable and unstable manifolds intersect transversally, i.e., the unstable manifold intersects and crosses the stable manifold. In two dimensions, continuous dynamical systems do not have transverse homoclinic orbits, but a two-dimensional Poincar\'e map defined near a periodic orbit of a continuous dynamical system may have them. 

\item Transverse homoclinic points obstruct the existence of second integrals of the motion. 
\item Transverse homoclinic points imply that chaotic motions exist nearby. 
\end{itemize}

The model problem, Eq. \ref{proxy}, is essentially a simple pendulum coupled weakly to a linear oscillator. For the restricted three body problem itself, 
Poincar\'e showed that after applying perturbation methods and truncating certain higher order terms in the expansion the Hamiltonian becomes completely integrable. He also showed that the reduced system, and therefore its Poincar\'e map, possesses hyperbolic saddle points whose stable and unstable manifolds, being level sets of the second integral, coincide, as they do for the pendulum illustrated in Figure \ref{pendulum}. He then asked the key question in the qualitative approach to dynamical systems: Should I expect this picture to persist if I restore the higher order terms? In other words, is the reduced system structurally stable? It is now known that integrable Hamiltonian systems of two or more degrees of freedom are not structurally stable. It is for this reason, even if no other, that they are exciting and productive to study.

In this section we have described how the structural stability of a Poincar\'e map of a continuous dynamical system can be evaluated, even though in general such a map cannot be computed explicitly. In the next section we look at stability and chaos in an explicit discrete dynamical system.

\section{Discrete dynamics, blowflies, feedback, and stability}\label{sec5}

In a series of population dynamics experiments, May and Oster and co-workers\cite{May:1976}
 chose to rear blowflies in boxes (for reasons we cannot entirely 
fathom --- surely there are more alluring model species), and count their numbers at every generation. 
The blowflies in their boxes are a simple ecological system  consisting of a single species limited by crowding and food supply, but with no predation. The system was analysed as a model of discrete chaos, and, in a different paradigm, as a control system by Mees (1981)\cite{Mees:1981}.

Assuming  discrete generations,  the data for the population dynamics of the blowflies can be fitted by a first-order difference equation \begin{equation}\label{e17}
N_{t+1}=f(N_t), 
\end{equation}
where $N$ is the number of blowflies in the time period $t$. The function $f$ is chosen so that $f(N_t)$ increases when the population is small, because
there is plenty of food and living space in the box,  but decreases when the population is large, because of competition for food and living space.       

The simplest single-humped function for $f$ that one can think of is a parabola:
\begin{equation}
f(N)= rN(1-N), \label{logistic}  
\end{equation}
for which Equation \ref{e17} is known as the logistic map. The parameter $r$ 
is then the reproduction rate constant. Equation \ref{e17} then says that due to reproduction the population will increase at a rate proportional to
 the current population, and due to starvation the population will 
decrease at a rate proportional to the \emph{square} of
 the current population. For example, if there is a large number of flies in a box in one time period, they will eat most of the food,
 and the next generation of flies will be few in number.

The weird properties of this simple model never fail to delight people. Their implications for ecologies were explored in May (1974)\cite{May:1974};  a good modern mathematical treatment, accompanied by downloadable software to play with, is given in Chapter~1 of Ball (2003)\cite{Ball-book:2003}. 

\begin{figure}[h]
\centerline{\hbox{
\includegraphics[scale=0.48]{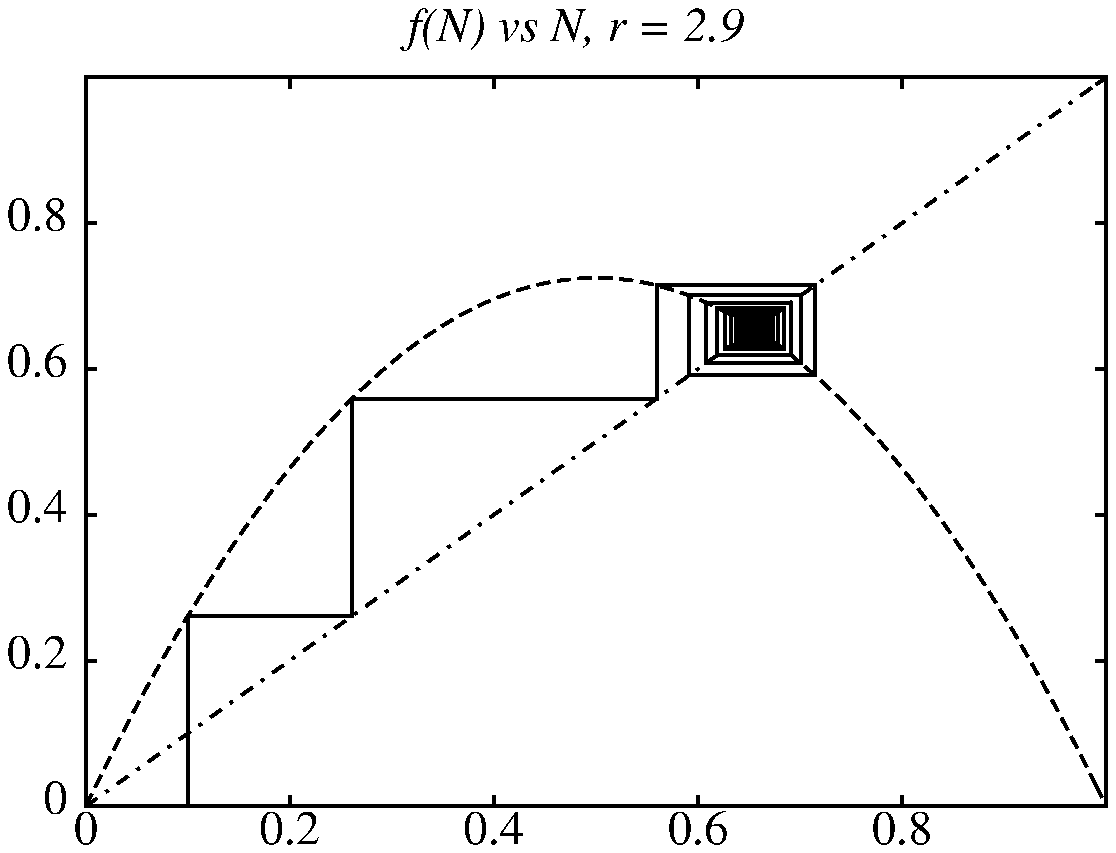}\hspace*{0.3cm}
\includegraphics[scale=0.48]{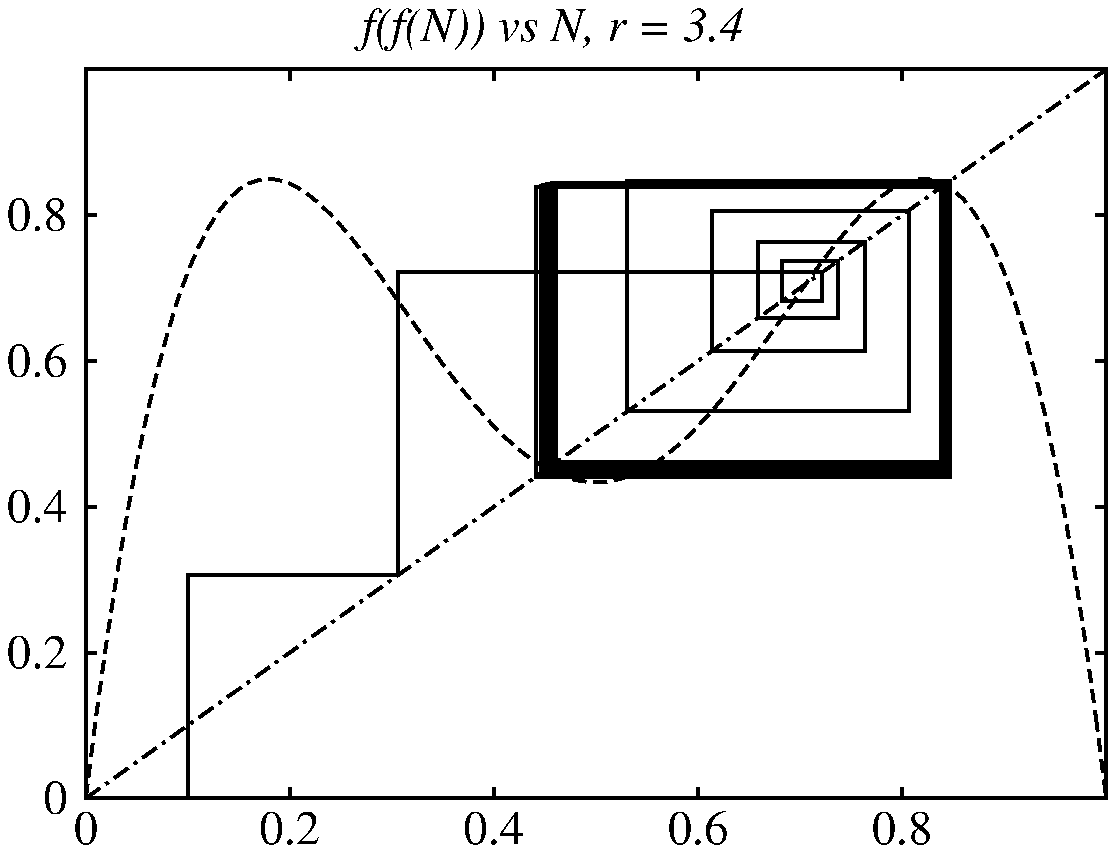}}}
\caption{\label{cobweb}  (left) The logistic function $f(N)$, Eq. \ref{logistic}, is plotted 
against the population $N$ for $r=2.9$,
  (right) the second composition $f(f(N))$
is plotted against $N$ for $r=3.4$. }
\end{figure}

The evolution of the population $N$ starting initially at $N_0$ 
may be found graphically as indicated in the \emph{cobweb diagram} of Figure \ref{cobweb}(a), where the $f(N)$ of Eq. \ref{logistic} is plotted against $N$ for a given value of $r$ (dashed curve). 
A vertical line takes the eye from $N_t$, the population in time-window $t$, to the corresponding $f(N_t)$ and 
an adjoining horizontal line takes you from $f(N_t)$ to $N_{t+1}$, the population in the time-window $t+1$. 
The solution converges to a point  of zero population growth where the graphs of  $f(N)= rN(1-N)$ and $f(N)= N$  intersect.  This period-1 fixed point (or equilibrium) is a stable attractor:
all nearby orbits converge to it as $t\rightarrow\infty$. 

Increasing the height of the hump, $r$, means increasing the reproduction rate in the blowfly model. For example, at 
$r=3.4$ the equilibrium has become  unstable and two new stable equilibria have appeared. These new equilibria are  not fixed points of $f$. They are fixed points of the second composition map,
$$f_2(N)\equiv f(f(N)),$$ 
 as shown in Figure  \ref{cobweb}(b). Here, the initial condition $N_0$ is the same as in (a), and the iterates at first take the population toward the old fixed point. But then they are repelled from it, because it is unstable, and converge instead to the \emph{two} intersections of
$f_2(N)=f(f(N))$ and $f_2(N)=N$, between which they oscillate in a 
period 2 orbit. 
This situation corresponds to the population $N$ switching between  two states: a highly populated generation results in the next generation being poorly populated, but then resources are plentiful enough to induce a populous generation again, and so on.

\begin{figure}
\centerline{\includegraphics[scale=1.2]{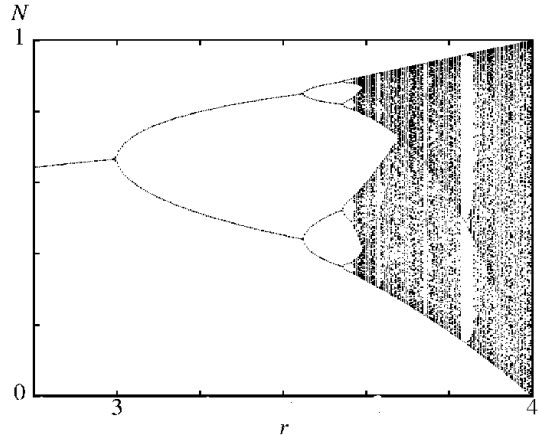}}
\caption{\label{bd} Bifurcation diagram over $r$ for the logistic map, where a point is plotted for each solution at every increment in $r$.}
\end{figure}

One cannot help but be curious as to what happens when the parameter~$r$ is increased again, and again \ldots We could compute many more of  these cobweb diagrams, each at a different value of $r$, but both the diagrams and this chapter would become very crowded. Our curiosity can be assuaged (or whetted!)  more succinctly by inspecting the bifurcation diagram of stable solutions 
in Figure \ref{bd}. 
One can easily make out the branch points at $r\approx
 3$, $3.449, \text{ and } 3.544$ 
corresponding to bifurcations to period 2, 4 and 8 orbits. Beyond that, the period-doubling repeats until the periodic behaviour of the population becomes chaotic. The population never settles to discernibly regular $n$-periodic oscillations, although the window at $r\sim 3.8$ suggests the resumption of some sort of regularity. 
 
\subsubsection*{Blowfly dynamics as a feedback system}
So far we have viewed the blowfly system as a difference equation, to model the generational delay, and as a bifurcation problem, to study the stability of the dynamics. Picking up the theme of section \ref{sec3}, it is also instructive to view the blowfly system as a simple feedback system. 

\begin{figure}\centerline{
\includegraphics[scale=0.6]{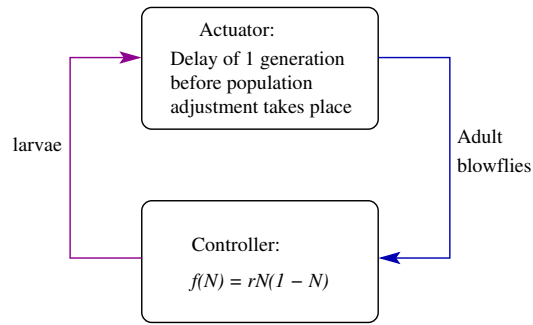}
}
\caption{\label{flies} Schematizing the blowfly system as a block diagram  brings out the feedback nature of the dynamics.}
\end{figure}

The output of the system (number of adult blowflies) is sensed by
a controller which implements a mechanism, approximated here by the model function $f(N)=rN(1-N)$ to control the level of input,
or number of larvae. The actuating mechanism which transforms the larvae into adult flies is simply the delay time of one generation. 
Figure \ref{flies} represents the feedback system as a block diagram. 

This diagram may seem rather facile, and nowhere near as interesting as the cobweb or bifurcation diagrams, but it does highlight a different side to the problem. For instance we see that the block components are independent. We could change the function $f(N)$ without changing the simple delay model. Inspecting this diagram also makes it easy to build in perturbations such as predation or injecting more flies from outside. 

The conceptual difference between modelling the blowfly population as a difference equation and as a feedback system is how information is treated. In the block diagram representation the information flow is explicit and the feedback is obvious, and
we can immediately think up ways of adding additional regulations to it. In this sense feedback is an \emph{information} science. This information  about connectivity is subsumed in the discrete dynamical model, which allows us to analyse the stability of the population but glosses over the fact that the instabilities are caused by feedback.

\section{ Stability of complex networks}\label{sec6}
The third (and final, for this chapter) destination in our world tour of nonlinear dynamics is the 41st floor of an office tower in the district of Wan Chai, Hong Kong. It is here that the transport operations and infrastructure of Hong Kong, Kowloon, and the New Territories (which together constitute a  Special Administrative Region of the People's Republic of China, or HKSAR) are controlled and coordinated day-to-day, and planning and policy development for future transport needs are carried out. 

The job of the HKSAR Department of Transport is formidable. Consider the problem: The public transport network carries over 11 million passenger trips \textit{each day} and this number will increase. It consists of railways, franchised buses, public light buses, private buses, ferries, trams, and taxis. Each of these components is a complex sub-network in its own right. The area is geographically diverse, with islands, harbour, waterways, steep hills, airport, and old built-up districts  with limited road space to be traversed or accessed. Environmental imperatives require the use of or conversion to low or zero emissions locomotive units. Efficient integration with transport in the densely populated economic-tiger zones of the Pearl River Delta is becoming necessary. The network as a whole must be safe, affordable, reliable, and robust. It must minimize redundancy and duplication of services, yet be flexible enough to match new demand without undue time-lags and provide services to new and changing population and employment centres. This means it must be capable of response and adaptatation on two time scales, daily and long-term (approximately yearly).
 
What a tall order! Can one tackle this complex network problem using the tools of dynamical systems theory? 
In dynamical systems language we ask: Is the HKSAR public transport network stable? Intuitively (or through direct experience) we expect such a complex network to exhibit \textbf{sensitive dependence on initial conditions}.  One blinking red LED on a signal-room console leads to a log-jam of peak hour trains. 
Even with no perturbations on the network itself we know (with depressing certitude) that leaving for work five minutes later than usual is likely to result in arriving at work an hour late.  These sorts of cascade effects in networks seem to occur  when a small disturbance in one element of a network is transmitted through it leading to instability as it spreads, but what lies behind these phenomena?

Studying networks such as the HKSAR public transport network is  about building models of how they function, and then analysing those models to understand how changes in the structure of the network will result in changes in behaviour.
Na\"ively, one expects that increasing the fraction of interacting elements or increasing the strength of interaction will enhance the  stability of a complex network, but as we will show in the next example, that is not necessarily so. 

In a paper in \textit{Nature} in 1972 Robert May\cite{May:1972} used  random matrix theory to show that in a large, linear, randomly coupled
network  the system dimension and the coupling strength must together satisfy a simple inequality.  Let us revisit the matrix equation (\ref{linear}):
\begin{equation}\tag{\ref{linear}}
\mathbf{\dot{x}}=\mathbf{A}\mathbf{x}. 
\end{equation}
May considered this as the linearization of a (large) set of nonlinear first-order differential equations that describe an ecology, or populations of $n$ interacting species, but it could equally well describe rates of passenger turnover at each of $n$ nodes in a public transport network.  The elements of the $n\times 1$ column vector $\mathbf{x}$ are the disturbed populations $x_j$ and the elements $a_{jk}$ of the $n\times n$ interaction matrix $\mathbf{A}$ describe the effect of species $k$ on species $j$ near equilibrium. Each $a_{jk}$ is assigned from a distribution of random numbers that has a mean of zero, so that any element is  equally likely to be positive or negative, and a mean square value $\alpha$, which expresses the average interaction strength. Then
$$\mathbf{A}= \mathbf{B}-\mathbf{I},$$
where $\mathbf{B}$ is a random matrix and $\mathbf{I}$ is the unit matrix. 

The probability that any pair of species will interact is expressed by the connectance $C$, measured as the fraction of non-zero elements in $\mathbf{A}$. The elements in the random matrix $\mathbf{B}$ are drawn from the random number distribution with probability $C$ or are zero with probability $1-C$. For any given system of size $n$, average interaction strength $\alpha$, and connectance $C$ we ask what is the probability $P(n,\alpha,C)$ that any particular matrix drawn from the ensemble gives a  stable system? May found that for large $n$ the system (\ref{linear}) is almost certainly stable ($P(n,\alpha,C\rightarrow 1)$) if
$$\alpha <(nC)^{-1/2} , $$
and almost certainly unstable ($P(n,\alpha,C\rightarrow 0)$) if
$$\alpha >(nC)^{-1/2} . $$
This result suggests that an ecology that is too richly connected (large $C$) or too strongly connected (large $\alpha$) is likely to be unstable and that the effect is more dramatic the larger the number of species $n$. 

May's result is based firmly on stability theory as it was developed by Poincar\'e and Lyapunov over a hundred years ago, as are more recent results on stability and control of dynamical network systems. For example, Yao \textit{et al}\cite{Yao:2005} in proposing a control method for chaotic systems with disturbances and unknown parameters (imprecisely modelled or unmodelled dynamics) rely on Lyapunov stability theory, as do  almost all of the applications mentioned by Boccaletti and Pecora (2006)\cite{Boccaletti:2006} in the preface to a special issue of the journal \textit{Chaos} devoted to stability of complex networks.

\section{Conclusions and inconclusions}
Although dynamical systems and stability theory was born and bred in celestial mechanics and control engineering, we now see that the concepts and methods have much wider application in the biological and environmental sciences and in socio-economic modelling and forecasting. 
A goal that is shared by many researchers in both hard and soft science is the improved management, and ultimately a priori design, of complex dynamical networks that are intrinsically imprecise or error-prone. To this end there is a need to disseminate the principles of stability and chaos outside mathematics, so that non-mathematical scientists are better-equipped to understand and manage the dynamics of complex natural and anthropogenic systems, and channel uncertainty into  stable output.


 How will these problems, fundamental and applied, be tackled? How will the science of dynamical systems, stability and chaos advance? We suggest that the three main  approaches will be used in synergy: qualitative and asymptotic analysis, 
interdisciplinary collaboration,
and computation.

The rapid growth of interest in dynamical systems and chaos over the past 30 years is, in a sense, quite different from the way that  areas of mathematics and physics developed in earlier times. It is not driven by industrialization, as for example was thermodynamics in the 19th century and classical control in the early 20th century, or by defence and cold war imperatives, as  was nuclear physics from the 1940s to the 1960s. 
What we are seeing now is the reverse: theory and mathematics of dynamical systems  and chaos together with faster computers are actually driving developments in a wide range of very 
diverse fields, from medical imaging to art restoration,  traffic control  to ecosystems,  neuroscience to climatology. 

\bibliographystyle{unsrt}


\newpage

\section*{Glossary}
The terms highlighted in bold-faced type in their first appearance in the  text are defined or described in this glossary. More comprehensive glossaries of dynamical systems terminology may be found easily on the web; for example, mrb.niddk.nih.gov/glossary/glossary.html,  www.dynamicalsystems.org/gl/gl/.    

\begin{description}
\item[Asymptotic solutions:] Solutions which asymptotically approach an unstable periodic solution. 
\item [Homoclinic points or doubly asymptotic solutions:] Points at which stable and unstable manifolds intersect transversally. In a Hamiltonian flow the stable and unstable manifolds must intersect transversally infinitely often (or coincide, as in the harmonic oscillator, Equation \ref{hpem}) because otherwise one of them would shrink and volume conservation would be violated. This remains true for dissipative systems\cite{Guck:1983}.
\item[Homoclinic chaos] or \textbf{homoclinic tangle} or \textbf{sensitive dependence on initial conditions}: A region densely packed with homoclinic points, where the dynamics is equivalent to and described by the Smale horseshoe map. Arbitrarily close initial conditions must actually belong to totally different parts of the homoclinic tangle,
 therefore they evolve quite differently in  time.  
\item[Poincar\'e map and cross section:] A sort of stroboscopic map; an extremely useful way of representing the dynamics of a two degree of freedom system on a plane. Consider the set of trajectories of a two degree of freedom Hamiltonian system that satisfy $H(p_1,p_2,q_1,q_2)=C$, where $C$ is a constant and $p_1,q_1$ and $p_2,q_2$ are canonical action-angle variables. Each energy level $H = C$ is therefore three-dimensional. To construct a Poincar\'e map we take a two-dimensional transverse surface or \textit{cross section} $\Sigma$ such as that defined by $q_2$=0. Then, for given $C$ the value of $p_2$ can be computed by solving the implicit equation $H(p_1,p_2,q_1,0)=C$, so that we may locally describe $\Sigma$ by the two variables $(q_1, p_1)$. Successive punctures of the surface $\Sigma$ in one direction by  each trajectory form a stroboscopic map of the time evolution of the trajectory in phase space.   
\item[Recurrence theorem:] A volume-preserving system has an infinite number of solutions which  return infinitely often to their initial positions, or an infinite number of Poisson stable solutions. 
\item [Hopf bifurcation:] The real parts of a pair of conjugate eigenvalues become positive and a family of periodic orbits bifurcates from a ``spiral'' fixed point (a focus). 
\item[Neimark-Sacker bifurcation] or secondary Hopf bifurcation: Consider a periodic orbit with period $T = 2\pi /\omega_1$ and suppose that a pair of Floquet multipliers crosses the unit circle at $\pm e^{i\omega_2}$ at an isolated bifurcation point. An invariant torus is born. Solutions on the torus are quasi-periodic, and if $q \omega_1 = p\omega_2$ for integers $p$ and $q$ the motion is said to be phase-locked. The Floquet multipliers are related to the eigenvalues of the Poincar\'e map linearised at the fixed point corresponding to the original $T$-periodic orbit.
\end{description}
\end{document}